\documentclass[10pt,conference,letterpaper]{IEEEtran}

\newtheorem{theorem}{Theorem}

\usepackage{booktabs}
\usepackage{bm}
\usepackage{graphicx}
\usepackage[cmex10]{amsmath}
\usepackage{cite}
\usepackage{amssymb}
\usepackage{subfigure}
\usepackage{extarrows}
\usepackage[letterspace = -2]{microtype}
\usepackage{algorithm}
\usepackage{algpseudocode}
\usepackage{algpascal}
\usepackage{float}

\usepackage[absolute,showboxes]{textpos}

\newfloat{algorithm}{t}{lop}

\TPMargin{5pt}

\newcommand{\copyrightstatement}{
    \begin{textblock}{0.84}(0.08,0.93)    
         \noindent
         \footnotesize
         $\copyright$ 2016 IEEE. Personal use of this material is permitted. Permission from IEEE must be obtained for all other uses, in any current or future media, including reprinting/republishing this material for advertising or promotional purposes, creating new collective works, for resale or redistribution to servers or lists, or reuse of any copyrighted component of this work in other works.
    \end{textblock}
}

\begin{document}
\copyrightstatement
\title{Waveform Optimization for Large-Scale Multi-Antenna Multi-Sine Wireless Power Transfer
}

\author{
\IEEEauthorblockN{Yang Huang\IEEEauthorrefmark{1} and Bruno Clerckx\IEEEauthorrefmark{1}\IEEEauthorrefmark{2}}

\IEEEauthorblockA{\IEEEauthorrefmark{1} Department of Electrical and Electronic Engineering, Imperial College London, London SW7 2AZ, United Kingdom}
\IEEEauthorblockA{\IEEEauthorrefmark{2} School of Electrical Engineering, Korea University, Korea}
\IEEEauthorblockA{Email: \{y.huang13, b.clerckx\}@imperial.ac.uk}
}

\maketitle

\vspace{-0.2cm}
\begin{abstract}
Wireless power transfer (WPT) is expected to be a technology reshaping the landscape of low-power applications such as the Internet of Things, machine-to-machine communications and radio frequency identification networks. Although there has been some progress towards multi-antenna multi-sine WPT design, the large-scale design of WPT, reminiscent of massive multiple-input multiple-output (MIMO) in communications, remains an open problem. Considering the nonlinear rectifier model, a multiuser waveform optimization algorithm is derived based on successive convex approximation (SCA). A lower-complexity algorithm is derived based on asymptotic analysis and sequential approximation (SA). It is shown that the difference between the average output voltage achieved by the two algorithms can be negligible provided the number of antennas is large enough. The performance gain of the nonlinear model based design over the linear model based design can be large, in the presence of a large number of tones\footnote{The work of Y. Huang was supported by CSC Imperial Scholarship.}.
\end{abstract}


\section{Introduction}
Wireless power transfer (WPT) technology is expected to be beneficial to the Internet of Things, machine-to-machine communications and radio frequency identification networks, thanks to the fully controlled power delivery. This paper focuses on the far-field WPT, where a rectenna is exploited to convert the electromagnetic radiation energy transmitted over long distance into DC power which can be stored in batteries\cite{BBFCGC15}. Improving the energy transfer efficiency is a key issue.

It is recently found that the efficiency is a function of the input waveforms, and the efficiency can be significantly improved by multi-sine signals\cite{BBFCGC15}. The question then arises as how to optimally design multi-sine waveform for WPT. In order to answer the question, the first issue to be tackled is the modeling of the nonlinear rectifying process. Although most off-the-shelf rectifier models in the context of microwave theory provide insights into the accurate rectifying process, the non-closed forms or highly complex structures in these models\cite{LW15, VMD15} make it hard to derive efficient algorithms for wireless transmissions. In contrast, to balance the accuracy and complexity in signal processing, \cite{CBYM15} constructs the model by truncating the Taylor expansion of the Shockley diode equation to the 4th order, as the 4th order truncation can describe the basic rectifying process\cite{LW15, BC11}. Based on this model, the waveform optimization problem is solved in \cite{CBYM15} (and further extended in \cite{CBmar16arxiv}) by reversed geometric programming (GP). This work confirms the adopted model by circuit simulation. The numerical results highlight the significant gains of the optimal waveforms over other waveforms as the number of sinewaves increases. This sheds interest on a large-scale design for WPT. Unfortunately, as reversed GP may take exponential time to compute the solution\cite{MChiang05}, the approach is highly complex for large-scale designs, although it can be extended to designs with higher-order truncation models.

Reminiscent of massive multiple-input multiple-output in communications, we investigate a multiuser large-scale multi-antenna multi-sine WPT. As \cite{CBYM15} and \cite{CBmar16arxiv} have shown that the 4\,th order truncation yields promising results, we also model the rectifier as a power series truncated to the 4th order. To avoid reversed GP, the model is finally reformulated as a scalar function of vector variables. Although the modeling method is motivated by \cite{Wetenkamp83} and different from the methods in \cite{CBYM15} and \cite{CBmar16arxiv}, the obtained model is equivalent to those in \cite{CBYM15} and \cite{CBmar16arxiv} with respect to (w.r.t.) optimization. To optimize the waveform, an algorithm is proposed based on successive convex approximation (SCA)\cite{MW78}. Then, a lower-complexity algorithm is proposed based on sequential approximation (SA), motivated by the law of large numbers derived from the independent and identically distributed (i.i.d.) circularly symmetric complex Gaussian (CSCG) random frequency/spatial domain channel gains. In the two algorithms, the approximate problem (AP) in each iteration yields a closed-form solution, which finally converges to a stationary point of its own original problem. It is shown that as the number of antennas increases, the output voltage maximized by the SA-based algorithm can be close to that offered by SCA. It is also shown that the average output voltage gain of the design based on the 4th order truncation model over the design based on the conventional linear model\footnote{In the conventional linear model, the harvested energy is a linear function of the average input power to the rectifier, which essentially is a 2nd order truncation model\cite{ZZH13}.} can be significantly high, because of the large number of tones.

\emph{Organizations}: The system model is elaborated in Section \ref{SecSystemModel}. Section \ref{SecWaveOptAlgos} proposes the waveform optimization algorithms. Section \ref{SecSimResults} discusses the simulation results. Conclusions are drawn in Section \ref{SecConclu}. \emph{Notations}: Matrices and vectors are in bold capital and bold lower cases, respectively. The notations {\small$(\cdot)^T$}, {\small$(\cdot)^\star$}, {\small$(\cdot)^\ast$}, {\small$(\cdot)^H$}, {\small$\text{Tr}\{\cdot\}$}, {\small$\|\cdot\|$}, and {\small$|\cdot|$} represent the transpose, optimal solution, conjugate, conjugate transpose, trace, 2-norm, and absolute value, respectively. The notation {\small$\mathbf{A} \succeq 0$} means that {\small$\mathbf{A}$} is positive-semidefinite.

\section{System Model}
\label{SecSystemModel}
\subsection{Signal Transmission}
\label{SecSignalTrans}
In the WPT system, a $M$-antenna base station (BS) delivers multi-sine energy signals over $N$ frequencies to $K$ single-antenna users. It is assumed that perfect channel state information (CSI) is available at the BS. All channel frequency responses remain constant during the transmission. The complex scalar frequency response of the channel between the $m$\,th antenna and the user $q$ (for {\small$q = 1,\ldots,K$}) at the $n$\,th frequency is designated as {\small$h_{q,(n-1)M + m}$} (for {\small$n = 1, \ldots, N$} and {\small$m = 1, \ldots, M$}), which is collected into {\small$\mathbf{h}_q \in \mathbb{C}^{MN \times 1}$}. Hence, {\small$\mathbf{h}_q = [\mathbf{h}_{q,1}^T, \ldots,$ $\mathbf{h}_{q,N}^T]^T$}, where {\small$\mathbf{h}_{q,n} = [h_{q,(n-1)M+1}, \ldots, h_{q,(n-1)M+M}]^T$} describes the spatial domain channel gains at the $n$\,th frequency.

The complex version of the transmitted signal at the $m$\,th BS antenna is {\small$\tilde{x}_m(t) = \sum_{n = 1}^N s_{(n-1)M+m} e^{j\omega_n t}$}, where the complex variable $s_{(n-1)M+m}$ collects the magnitude and the initial phase of the radio-frequency (RF) complex signal at angular frequency $\omega_n$. Hence, the RF signal transmitted by antenna $m$ is {\small $x_m(t) \triangleq \sqrt{2} \text{Re}\{\tilde{x}_m(t)\}$}. The variable $s_{(n-1)M+m}$ is collected into {\small$\mathbf{s} \in \mathbb{C}^{MN \times 1}$}, such that $\mathbf{s} = [\mathbf{s}_1^T, \ldots, \mathbf{s}_N^T]^T$ and $\mathbf{s}_n = [s_{(n-1)M+1}, \ldots, s_{(n-1)M+M}]^T$ describes all the signals transmitted at angular frequency $\omega_n$, where {\small$\omega_n \!=\! \omega_1 \! + \! (n\!-\!1)\Delta_\omega$}, for {\small$n \! = \! 1, \ldots, N$} and {\small$\omega_1 \!> \! (N \! - \! 1)\Delta_\omega/2$}. Suppose the BS transmit power is constrained by $P$, such that {\small $\|\mathbf{s}\|^2 = \sum_{n = 1}^N \sum_{m = 1}^M |s_{(n-1)M + m}|^2 \leq P$}. The complex RF signal through the channel between the $m$\,th transmit antenna and the $q$\,th user can be written as {\small $\tilde{y}_{q\!,m}\!(t) = \sum_{n = 1}^N s_{(n\!-\!1)M+m} h_{q,(n-1)M+m} e^{j\omega_n t}$}. Hence, the RF signal transmitted from the $M$ antennas and input into the antenna at user $q$ is given by {\small$y_q(t) = \sqrt{2} \text{Re}\big\{\sum_{m = 1}^M \tilde{y}_{q,m}(t)\big\}$}.

\subsection{Modeling the Nonlinear Rectifying Process}
\begin{figure}[!t]
\centering
\includegraphics[width = 3.0in]{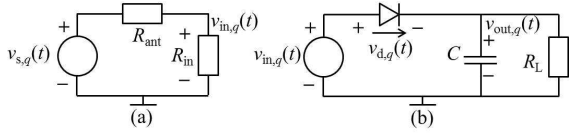}
\caption{Simplified rectifier circuit for analysis. The capacitor $C$ functions as a low-pass filter, and $R_{\text{L}}$ is the load.}
\label{FigEHcirAna}
\end{figure}
\begin{figure}[!t]
\centering
\includegraphics[width = 2.6in]{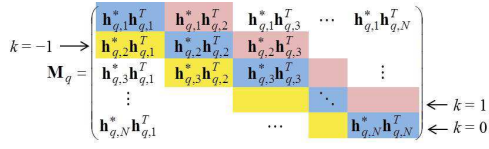}
\caption{$\mathbf{M}_{q,1}$ is the above matrix only maintaining the block diagonal (whose index is $k=1$) in pink, while all the other blocks are set as $\mathbf{0}_{M\times M}$.}
\label{FigM_mat}
\end{figure}
As shown in Fig. \ref{FigEHcirAna}(a), the antenna at user $q$ is modeled as a source {\small$v_{\text{s},q}(t)$} in series with an impedance {\small$R_{\text{ant}} = 50\Omega$}. Assuming a lossless antenna, all the input power to the rectenna can be absorbed by the rectifier's input impedance {\small$R_{\text{in}}$}, such that {\small$\mathcal{E}\{|v_{\text{in},q}(t)|^2\}/R_{\text{in}} = \mathcal{E}\{|y_q(t)|^2\}$}. Maximizing the power dissipated in {\small$R_{\text{in}}$} yields {\small$R_{\text{ant}} = R_{\text{in}}$}. Assuming an ideal matching network, the input voltage to the rectifier equals {\small$v_{\text{in},q}(t) = y_q(t)\sqrt{R_{\text{ant}}}$}, such that {\small$v_{\text{s},q}(t) = 2 y_q(t)\sqrt{R_{\text{ant}}}$}.

As shown in Fig. \ref{FigEHcirAna}(b), the input signal is rectified by a Schottky diode and goes through a low pass filter. Motivated by \cite{Wetenkamp83}, we model {\small$v_{\text{out},q}(t)$} as an approximate function of the input waveform, by manipulating the Shockley diode equation. The Shockley equation shows that the diode output current {\small$i_{\text{d},q}(t)\! =\! i_\text{s}(\exp(\frac{v_{\text{d},q}(t)}{nV_\text{T}})\! - \! 1)$}, where {\small$v_{\text{d},q}(t)\! =\! v_{\text{in},q}(t)\! -\! v_{\text{out},q}(t)$}, while {\small$i_\text{s}$}, {\small$V_\text{T}$} and {\small$n$} represent the saturation current, the thermal voltage and the ideality factor (set to 1 for simplicity), respectively. Due to the low-power input and the high impedance load, $i_{\text{d},q}(t)$ can be approximated as zero\cite{Wetenkamp83}. Ref. \cite{Wetenkamp83} essentially achieves a 2nd order truncation model, while the 4th-order truncation is necessary for describing the basic diode rectifying process\cite{LW15, BC11}. Therefore, to finally achieve the 4th order truncation model, the Taylor expansion of $e^x$ for {\small$x = \frac{v_{\text{in},q}(t)}{nV_\text{T}}$} is applied to the Shockley equation. Hence, {\small$v_{\text{out},q}(t) = n V_\text{T} \! \ln \! \Big[ \!1 \! + \! \frac{v_{\text{in},q}(t)}{n V_\text{T}} \! + \! \frac{v_{\text{in},q}^2(t)}{2 n^2 V_\text{T}^2} \! + \! \frac{v_{\text{in},q}^3(t)}{6 n^3 V_\text{T}^3} \! + \! \frac{v_{\text{in},q}^4(t)}{24 n^4 V_\text{T}^4}\Big]$}, where the Taylor series is truncated to the 4th order. By assuming an ideal low-pass filter, $v_{\text{out},q}(t)$ yields DC voltage, i.e. $v_{\text{out},q}(t) = v_{\text{out},q}$. Not contributing to the DC voltage, the odd order terms in the Taylor series can be omitted. Thus, {\small$v_{\text{out},q} = n V_\text{T} \ln \Big[1 + \frac{f_\text{LPF}(v_{\text{in},q}^2(t))}{2 n^2 V_\text{T}^2} + \frac{f_\text{LPF}(v_{\text{in},q}^4(t))}{24 n^4 V_\text{T}^4}\Big]$}, where $f_\text{LPF}(\cdot)$ represents the ideal low-pass filtering process, which omits the non-DC harmonics. Making use of $\ln(1 + x) \simeq x$ yields {\small$v_{\text{out},q} = \beta_2 f_\text{LPF}\big(y_q^2(t)\big) + \beta_4 f_\text{LPF}\big(y_q^4(t)\big)$}, where {\small$\beta_2 = R_\text{ant}/(2 n V_\text{T})$} and {\small$\beta_4 = R_\text{ant}^2/(24 n^3 V_\text{T}^3)$}. It is noticeable that multiplying the above {\small$v_{\text{out},q}$} by {\small$i_\text{s}/(nV_\text{T})$} achieves nothing but the model {\small$z_{DC}$} in \cite{CBYM15}. In the term {\small$f_{\text{LPF}}\left(y_q^2(t)\right) = f_{\text{LPF}}\big(\text{Re}\{ \tilde{y}_q(t)\tilde{y}_q(t) \! + \tilde{y}_q(t)\tilde{y}_q^\ast(t) \}\big)$}, {\small$\text{Re}\{\tilde{y}_q(t)\tilde{y}_q(t)\}$} can be omitted as it only contains the non-DC harmonics. Hence, {\small$f_{\text{LPF}} \left(y_q^2(t)\right) = f_{\text{LPF}}\big(\text{Re}\{ \tilde{y}_q(t)\tilde{y}_q^\ast(t) \}\big) = \text{Re}\big\{\! \sum_{n \!=\! 1}^N \! \sum_{m_1\!,m_2}\! s_{(n\!-\!1)\!M\!+\!m_1} h_{q,(n\!-\!1)\!M\!+\!m_1} s^\ast_{(n\!-\!1)\!M \! + \!m_2} h^\ast_{q,(n\!-\!1)\!M\!+\!m_2}\! \big\}$ $= \sum_{n = 1}^N \mathbf{s}_n^H \mathbf{h}_{q,n}^\ast \mathbf{h}_{q,n}^T \mathbf{s}_n$}, where {\small$m_1, m_2 \in \{1, \ldots, M\}$}. Similarly, {\small$f_\text{LPF}\!\left(y_q^4(t)\right)\! =\! \frac{3}{2} \! \sum_{\substack{{n_1, n_2, n_3, n_4}\\{n_1 \! - \!  n_3 \! = \! - \! (n_2 \! - \! n_4)}}} \! \mathbf{s}_{\!n_3}^H \mathbf{h}_{q\!, n_3}^\ast  \mathbf{h}_{q\!, n_1}^T \mathbf{s}_{\!n_1} \mathbf{s}_{\!n_4}^H  \mathbf{h}_{q\!,n_4}^\ast \mathbf{h}_{\!n_2}^T \mathbf{s}_{q\!,n_2}$}, where $n_1, n_2, n_3, n_4 \in \{1, \dots, N\}$. It is inferred that if higher order truncations were considered, {\small$v_{\text{out},q}$} could not be written as the above vector formulation, such that the output voltage maximization problem in a form of polynomials has to be solved by the complex reversed GP algorithm in \cite{CBYM15}.

So far, it has been shown that $v_{\text{out},q}$ can be modeled as a function of vector variables $\{\mathbf{s}_n\}_{n=1}^N$. Fortunately, this function can be homogenized, by introducing {\small$MN$}-by-{\small$MN$} matrices {\small$\mathbf{M}_q \! \triangleq \! \mathbf{h}_q^\ast \mathbf{h}_q^T$} and {\small$\mathbf{M}_{q,k}$}.
As shown in Fig. \ref{FigM_mat}, {\small$k \! \in \! \{1, \ldots, N\!-\!1\}$} is the index of the $k$\,th block diagonal above the main block diagonal (whose index $k=0$) of {\small$\mathbf{M}_q$}, while {\small$k \! \in \! \{-\!(N\!-\!1), \ldots, -1\}$} is the index of the $|k|$\,th block diagonal below the main block diagonal. Given a certain $k$, {\small$\mathbf{M}_{q,k}$} is generated by retaining the $k$\,th block diagonal of {\small$\mathbf{M}_q$} but setting all the other blocks as {\small$\mathbf{0}_{M\times M}$}. For $k\geq 1$, the non-Hermitian matrix {\small$\mathbf{M}_{q,-k} = \mathbf{M}_{q,k}^H$}, while {\small$\mathbf{M}_{q,0} \succeq 0$}. Hence, $v_{\text{out},q}$ is finally formulated as
\begin{IEEEeqnarray}{rcl}
\label{EqFuncVoutq}
v_{\text{out},q} & {}={} & \beta_2 \mathbf{s}^H \mathbf{M}_{q,0} \mathbf{s} \! + \! \frac{3}{2}\beta_4 \mathbf{s}^H \mathbf{M}_{q,0} \mathbf{s}\!\left(\mathbf{s}^H \mathbf{M}_{q,0}\mathbf{s}\right)^H \! + \nonumber\\
&&\! 3\beta_4 \textstyle{\sum_{k = 1}^{N-1}} \mathbf{s}^H \mathbf{M}_{q,k} \mathbf{s} \!\left(\mathbf{s}^H \mathbf{M}_{q,k} \mathbf{s}\right)^H\,. \label{Eqv_outHomo}
\end{IEEEeqnarray}

\section{Waveform Optimization Algorithm}
\label{SecWaveOptAlgos}
\subsection{Waveform Optimization Based on SCA}
\label{SecGeneWirelessChan}
In order to address the waveform design problem of a $K$-user system, this section proposes an efficient algorithm based on solving the weighted-sum output voltage maximization problem given by $\max_{\mathbf{s}} \big\{ \sum_{q=1}^K w_q \!\cdot\! v_{\text{out},q}\big(\mathbf{s}\big)\!:\! \|\mathbf{s}\|^2 \!\leq\! P  \big\}$, where $w_q \geq 0$ $\forall q$ represent the users' weights. The problem can be equivalently formulated in its epigraph form:
\begin{IEEEeqnarray}{cl}
\min_{\gamma_1,\,  \mathbf{s}} \quad & \gamma_1 \IEEEyesnumber\IEEEyessubnumber\\
\text{s.t.} \,& - \textstyle{\sum_{q=1}^K} \!w_q \cdot v_{\text{out},q}\left(\mathbf{s}\right)  - \gamma_1 \leq 0 \,, \IEEEyessubnumber \label{EqHighDegPolyConst}\\
& \mathbf{s}^H\mathbf{s} \leq P\,. \IEEEyessubnumber
\end{IEEEeqnarray}
In order to make the above problem tractable, auxiliary variables $t_{q,k}$ (for {\small$k = 0, \ldots, N\!-\!1$}) are introduced, such that {\small$\mathbf{s}^H \mathbf{M}_{q,k} \mathbf{s} = \text{Tr}\{\mathbf{M}_{q,k} \mathbf{s}\mathbf{s}^H\} = \text{Tr}\{\mathbf{M}_{q,k} \mathbf{X}\} = t_{q,k}$}.
Therefore, the problem can be equivalently reformulated as
\begin{IEEEeqnarray}{cl}
\label{Epi_Problem_MaxVout}
\min_{\gamma_1,\, \{\mathbf{t}_q\}_{q=1}^K,\, \mathbf{X}\succeq 0} \, & \gamma_1 \IEEEyesnumber\IEEEyessubnumber\\
\text{s.t.} \,& \textstyle{\sum_{q=1}^K} w_q \! \left(- \beta_2 t_{q,0} + \mathbf{t}_q^H \mathbf{A}_0 \mathbf{t}_q\right) \! - \! \gamma_1 \! \leq \! 0, \IEEEyessubnumber \label{Eq_MaxVout_RankConstRelaxed_nonCVXquadrConst}\\
& \text{Tr}\{\mathbf{M}_{q,k} \mathbf{X}\} = t_{q,k}\,,\quad \forall q, k\,, \IEEEyessubnumber \label{EqConst_t_qk}\\
& \text{Tr}\{\mathbf{M}_{q,k}^H \mathbf{X}\} = t_{q,k}^\ast\,,\quad \forall q, k\neq0\,, \IEEEyessubnumber \label{EqConstConjugatet_qk}\\
& \text{Tr}\{\mathbf{X}\} \leq P\,, \IEEEyessubnumber \label{EqTxPwrConst}\\
& \text{rank}\{\mathbf{X}\} = 1\,,\IEEEyessubnumber \label{EqEpiMaxVoutRankConst}
\end{IEEEeqnarray}
where $\mathbf{A}_0 =  diag\{-\frac{3}{2}\beta_4, -3\beta_4, \ldots, -3\beta_4\}  \preceq  0$ and $\mathbf{t}_q = [t_{q,0},\ldots,t_{q,N-1}]^T$, such that $g_q(\mathbf{t}_q) \triangleq \mathbf{t}_q^H \mathbf{A}_0 \mathbf{t}_q = - \frac{3}{2} \beta_4 t_{q,0} t_{q,0}^\ast - 3\beta_4 \sum_{k=1}^{N-1}t_{q,k} t_{q,k}^\ast$. To make problem (\ref{Epi_Problem_MaxVout}) more tractable, relaxing the nonconvex rank constraint (\ref{EqEpiMaxVoutRankConst}) yields
\begin{IEEEeqnarray}{cl}
\label{Epi_Problem_MaxVout_RankConstRelaxed}
\min_{\gamma_1,\, \{\mathbf{t}_q\}_{q=1}^K,\, \mathbf{X}\succeq 0} \quad & \left\{ \gamma_1: \text{(\ref{Eq_MaxVout_RankConstRelaxed_nonCVXquadrConst}), (\ref{EqConst_t_qk}), (\ref{EqConstConjugatet_qk}), and (\ref{EqTxPwrConst})} \right\}\,.
\end{IEEEeqnarray}
Problem (\ref{Epi_Problem_MaxVout_RankConstRelaxed}) is still nonconvex, due to the nonconvex quadratic function $g_q(\mathbf{t}_q)$ in (\ref{Eq_MaxVout_RankConstRelaxed_nonCVXquadrConst}). Thus, SCA can be exploited to solve (\ref{Epi_Problem_MaxVout_RankConstRelaxed}). However, the solution of (\ref{Epi_Problem_MaxVout_RankConstRelaxed}) may be an infeasible solution of the original problem (\ref{Epi_Problem_MaxVout}), due to the rank relaxation. Fortunately, it is then shown that the solution of (\ref{Epi_Problem_MaxVout_RankConstRelaxed}) can satisfy the rank-1 constraint in (\ref{Epi_Problem_MaxVout}).


\subsubsection{Successive Convex Approximation}
\label{SecWeighSumSCAbasedAlg}
Problem (\ref{Epi_Problem_MaxVout_RankConstRelaxed}) is then approximated iteratively by SCA.
The nonconvex $g_q(\mathbf{t}_q)$ is approximated (at a certain point $\mathbf{\hat{t}}_q$) as a linear function by its first-order Taylor expansion $\tilde{g}_q(\mathbf{t}_q; \mathbf{\hat{t}}_q) \triangleq \mathbf{\hat{t}}_q^H \! \mathbf{A}_0 \! \mathbf{t}_q + \mathbf{\hat{t}}_q^T \! \mathbf{A}_0^T \! \mathbf{t}_q^\ast - \mathbf{\hat{t}}_q^H \! \mathbf{A}_0 \! \mathbf{\hat{t}}_q = 2\text{Re}\{\mathbf{\hat{t}}_q^H \mathbf{A}_0 \mathbf{t}_q\} - \mathbf{\hat{t}}_q^H \mathbf{A}_0 \mathbf{\hat{t}}_q$. Note that $g_q(\mathbf{t}_q) \leq \tilde{g}_q(\mathbf{t}_q; \mathbf{\hat{t}}_q)$, as $-g_q(\mathbf{t}_q)$ is convex. Suppose $\mathbf{t}_q^{(l - 1)}$ as the optimal $\mathbf{t}_q^\star$ approximated at iteration $(l - 1)$. Then, $\mathbf{t}_q^{(l - 1)}$ can be involved in the approximation in the next iteration $l$, by approximating $g_q(\mathbf{t}_q)$ as $\tilde{g}_q(\mathbf{t}_q; \mathbf{t}_q^{(l - 1)})$. Therefore, the $l$\,th convex AP can be formulated as
\begin{IEEEeqnarray}{cl}
\label{ApproxConvProblem_MaxVout}
\min_{\gamma_1\!,\{\!\mathbf{t}_q\!\}_{q\!=\!1}^K,\mathbf{X}\succeq 0} \, & \gamma_1 \IEEEyesnumber\IEEEyessubnumber \label{EqObjApproxCvxProb}\\
\text{s.t.} \,& \textstyle{\sum_{q = 1}^K} \! w_q \! \left(\! - \beta_2 t_{q,0} \! + \! \tilde{g}_q\!(\mathbf{t}_q; \mathbf{t}_q^{(l \! - \! 1)})\!\right) \! - \! \gamma_1 \! \leq \! 0, \IEEEyessubnumber \label{EqApproxCvxProb_LinearizedConst}\\
& \text{(\ref{EqConst_t_qk}), (\ref{EqConstConjugatet_qk}), and (\ref{EqTxPwrConst})}\,,\nonumber
\end{IEEEeqnarray}

\subsubsection{Solving the Approximate Convex Problem}
\label{SecSolveApproxProb}
The following Theorem \ref{TheoSCAconvergeOrigProb} shows that the semidefinite problem (SDP) (\ref{ApproxConvProblem_MaxVout}) can yield an optimal $\mathbf{X}^\star$ of rank 1, which means that when the solution of (\ref{ApproxConvProblem_MaxVout}) converges over iterations and remains rank-1, the final solution can be the solution of (\ref{Epi_Problem_MaxVout}).
\begin{theorem}
\label{TheoSCAconvergeOrigProb}
Problem (\ref{ApproxConvProblem_MaxVout}) has, among others, an optimal solution with a rank-1 $\mathbf{X}^\star$.
\end{theorem}
\begin{IEEEproof}
Substituting (\ref{EqConst_t_qk}) and (\ref{EqConstConjugatet_qk}) into (\ref{EqApproxCvxProb_LinearizedConst}) shows that problem (\ref{ApproxConvProblem_MaxVout}) essentially is an equivalent form of
\begin{IEEEeqnarray}{cl}
\label{ApproxConvProblem_MaxVoutEquiv}
\min_{\mathbf{X}\succeq 0} \, & \left\{\text{Tr}\{\mathbf{A}_1\mathbf{X}\}: \text{Tr}\{\mathbf{X}\} \leq P \right\}\,,
\end{IEEEeqnarray}
where $\mathbf{A}_1 \triangleq \mathbf{C}_1 + \mathbf{C}_1^H$ is Hermitian, and {\small$\mathbf{C}_1 \! = \! \sum_{q = 1}^K \! w_q \!\big(\! -\frac{\beta_2 \! + \! 3\beta_4t^{(l\!-\!1)}_{q,0}}{2}\mathbf{M}_{q,0} \! - \! 3\beta_4 \! \sum_{k=1}^{N\!-\!1} \![t^{(l \!- \!1)}_{q,k}]^\ast \!\mathbf{M}_{q,k} \! \big)$}. Proposition 3.5 in \cite{HP10} shows that problem (\ref{ApproxConvProblem_MaxVoutEquiv}) has, among others, a rank-1 solution. Because of the equivalence, the optimal solution of (\ref{ApproxConvProblem_MaxVoutEquiv}) also satisfies the KKT conditions of (\ref{ApproxConvProblem_MaxVout}). As (\ref{ApproxConvProblem_MaxVout}) is convex, the solution is the global optimum of (\ref{ApproxConvProblem_MaxVout}).
\end{IEEEproof}

In order to obtain a rank-1 solution $\mathbf{X}^\star$ in (\ref{ApproxConvProblem_MaxVout}), if we solve the SDP (\ref{ApproxConvProblem_MaxVoutEquiv}) by CVX with the interior point method \cite{GB14} and obtain the rank-1 solution by rank reduction\cite{HP10}, the complexity of solving the SDP is {\small$O(1)(2+2MN)^{1/2}(MN)^2\big(5(MN)^4 + 8(MN)^3 + (MN)^2 +1\big)$}\cite{BN01}. Fortunately, the following method yields a closed-form solution, with reduced complexity.

Given that problem (\ref{ApproxConvProblem_MaxVoutEquiv}) yields a rank-1 solution $\mathbf{X}^\star = \mathbf{x}^\star[\mathbf{x}^\star]^H$, (\ref{ApproxConvProblem_MaxVoutEquiv}) is equivalent to a nonconvex quadratically constrained quadratic problem (QCQP) given by
\begin{IEEEeqnarray}{cl}
\label{EquiQCQPApproxConvProblem}
\min_{\mathbf{x}} \quad & \left\{ \mathbf{x}^H\mathbf{A}_1\mathbf{x}: \|\mathbf{x}\|^2 \leq P \right\}\,.
\end{IEEEeqnarray}
Analyzing the KKT conditions shows that if {\small$\mathbf{A}_1 \succeq 0$} or {\small$\mathbf{A}_1 \succ 0$}, the optimal {\small$\mathbf{x}^\star = \mathbf{0}_{MN\times 1}$}. Otherwise, given the eigenvectors {\small$\mathbf{U}_{\mathbf{A}_1}$} of {\small$\mathbf{A}_1$}, the optimal {\small$\mathbf{x}^\star = \sqrt{P}\left[\mathbf{U}_{\mathbf{A}_1}\right]_{\text{min}}$}, where {\small $\left[\mathbf{U}_{\mathbf{A}_1}\right]_{\text{min}}$} is the eigenvector corresponding to the minimum eigenvalue of {\small$\mathbf{A}_1$}. Performing eigenvalue decomposition (EVD) for {\small$\mathbf{A}_1$} by the QR algorithm yields complexity of {\small$O\big((MN)^3\big)$}\cite{Parlett00}.

{\small
\begin{algorithm}
\caption{SCA-based weighted sum maximization}\label{AlgSCA}
\begin{algorithmic}[1]
\State \textbf{Initialize} Set $l = 0$, and generate feasible initial points $\mathbf{X}^{(0)}$, $\{\mathbf{t}_q^{(0)}\}_{q = 1}^K$ and $\gamma_1^{(0)}$;
\Repeat
    \State $l = l + 1$;
    \State Compute $\mathbf{A}_1$;
    \State \textbf{if} $\mathbf{A}_1 \succeq 0$ or $\mathbf{A}_1 \succ 0$ \textbf{then} $\mathbf{x}^\star = \mathbf{0}$; $\mathbf{X}^\star = \mathbf{x}^\star[\mathbf{x}^\star]^H$;
    \State \textbf{else} $\mathbf{x}^\star = \sqrt{P}\left[\mathbf{U}_{\mathbf{A}_1}\right]_{\text{min}}$; $\mathbf{X}^\star = \mathbf{x}^\star[\mathbf{x}^\star]^H$;
    \State Update $\mathbf{X}^{(l)} = \mathbf{X}^\star$; Update $t_{q,k}^{(l)}$ $\forall q,k$ by (\ref{EqConst_t_qk});
\Until{\|\mathbf{X}^{(l)} - \mathbf{X}^{(l-1)}\|_F/\|\mathbf{X}^{(l)}\|_F \leq \epsilon}
\end{algorithmic}
\end{algorithm}
}
The overall algorithm is summarized in Algorithm \ref{AlgSCA}. Since {\small$\tilde{g}_q(\mathbf{t}_q^{(l)}; \mathbf{t}_q^{(l)}) = g_q(\mathbf{t}_q^{(l)}) \leq \tilde{g}_q(\mathbf{t}_q^{(l)}; \mathbf{t}_q^{(l-1)})$}, the optimal solution {\small$\mathbf{X}^{(l-1)}$} of the {\small$(l\!-\!1)$}\,th AP (\ref{ApproxConvProblem_MaxVout}) is a feasible point of the $l$\,th AP (\ref{ApproxConvProblem_MaxVout}). As the AP (\ref{ApproxConvProblem_MaxVout}) is convex, the objective function (\ref{EqObjApproxCvxProb}) converges. Then, it can be shown that {\small$\{\mathbf{X}^{(l)}\}_{l=0}^{\infty}$} is a convergent sequence. As {\small$\nabla g_q(\mathbf{t}_q^{(l)}) = \nabla \tilde{g}_q(\mathbf{t}_q^{(l)}; \mathbf{t}_q^{(l)})$}, the solution of (\ref{ApproxConvProblem_MaxVout}) finally converges to a stationary point of (\ref{Epi_Problem_MaxVout_RankConstRelaxed}). The rank-1 solution of (\ref{Epi_Problem_MaxVout_RankConstRelaxed}) is also the stationary point of (\ref{Epi_Problem_MaxVout}). Thus, Algorithm \ref{AlgSCA} converges to a stationary point of problem (\ref{Epi_Problem_MaxVout}). The detailed proof is omitted due to space constraint.

\subsection{Algorithm Based on Large-Scale Systems}
\label{SecAsympAna}
{\small
\begin{algorithm}
\caption{SA-based weighted sum maximization}\label{AlgAsymptWsum}
\begin{algorithmic}[1]
\State \textbf{Initialize} Set $l = 0$, and generate feasible initial points $\{\mathbf{p}_q^{(0)}\}_{q=1}^K$. Then, compute $\{\mathbf{t}_q^{(0)}\}_{q = 1}^K$ by (\ref{EqConst2_EquiProbAsymptWsum}).
\Repeat
    \State $l = l + 1$;
    \State Compute $\{\mathbf{C}_{q,1}^\prime(\mathbf{t}_q^{(l-1)})\}_{q=1}^K$ and $\{\mathbf{A}^\prime_{q,1}(\mathbf{t}_q^{(l-1)})\}_{q=1}^K$;
    \State Compute $\mathbf{A}^\prime_1$, $\mathbf{\bar{p}}^\star = b \big[ \mathbf{U}_{\mathbf{\bar{\Lambda}}} \big]_\text{min}$ and $\{\mathbf{p}_q^\star\}_{q=1}^K$;
    \State Update $\mathbf{\bar{p}}^{(l)} = \mathbf{\bar{p}}^\star$ and $\mathbf{p}_q^{(l)} = \mathbf{p}_q^\star, \forall q$; Update $t_{q,k}^{(l)}$;
\Until{\|\mathbf{\bar{p}}^{(l)} - \mathbf{\bar{p}}^{(l-1)}\|/\|\mathbf{\bar{p}}^{(l)}\| \leq \epsilon}
\end{algorithmic}
\end{algorithm}}
We assume that the channel of a given user is sufficiently frequency-selective such that channel gains can be i.i.d. in space and frequency, and each channel gain {\small$h_{q,(n-1)M+1} \! \sim \! \mathcal{CN}(0, \Lambda_q)$}, where {\small$\Lambda^{1/2}_q$} is the large-scale fading. Channels of different users are also assumed fully uncorrelated. Therefore, the law of large numbers can be applied. Namely, as {\small$M\! \rightarrow\! \infty$}, {\small$\mathbf{h}_{q,n}^T \mathbf{h}_{q,n}^\ast/M\! = \! \Lambda_q$} and {\small$\mathbf{h}_{q,n}^T \mathbf{h}_{q^\prime,n^\prime}^\ast/M\! =\! 0$} for {\small$q^\prime \!\neq\! q$} or {\small$n^\prime \!\neq \! n$}.

The normalized asymptotically optimal $\mathbf{s}$ (defined in Section \ref{SecSignalTrans}) is designated as {\small$\mathbf{\bar{s}} = [\mathbf{\bar{s}}_1^T, \ldots, \mathbf{\bar{s}}_N^T]^T$}, such that $\mathbf{\bar{s}}_n$ is subject to {\small$\sum_{n = 1}^N \|\mathbf{\bar{s}}_n\|^2 = 1$}. Then, the optimal structure of $\mathbf{\bar{s}}_n$ can be $\mathbf{\bar{s}}_n = \sum_{q = 1}^K \xi_{q,n}\mathbf{h}_{q,n}^\ast/\sqrt{M}$, where $\xi_{q,n}$ is a complex weight. With such a $\mathbf{\bar{s}}_n$, by defining {\small$E \triangleq PM$}, the asymptotically optimal $\mathbf{s}$ can be written as {\small$\mathbf{s}_\text{asym} \triangleq \sqrt{E/M}\mathbf{\bar{s}}$}. The optimality of $\mathbf{\bar{s}}_n$ can be shown by contradiction as in\cite{XTW14}. Substituting $\mathbf{s}_\text{asym}$ into (\ref{EqFuncVoutq}) and applying the law of large numbers, the asymptotic output voltage at user $q$ can be written as
\begin{IEEEeqnarray}{rcl}
\label{EqAsymptOutputVol}
v_{\text{out},q}^\prime &{}={}& \beta_2 E\Lambda_q^2\mathbf{p}_q^H\mathbf{p}_q \! + \! \frac{3}{2}\beta_4 E^2\Lambda_q^4 \! \left(\!\mathbf{p}_q^H \mathbf{M}_0^\prime \mathbf{p}_q \! \right)\!\left(\mathbf{p}_q^H \mathbf{M}_0^\prime \mathbf{p}_q \! \right)^H \nonumber\\
&&{}+{}3\beta_4 E^2\Lambda_q^4 \textstyle{\sum_{k=1}^{N-1}}\! \left(\!\mathbf{p}_q^H \mathbf{M}_k^\prime \mathbf{p}_q \! \right)\!\left(\mathbf{p}_q^H \mathbf{M}_k^\prime \mathbf{p}_q \! \right)^H\,,
\end{IEEEeqnarray}
where {\small$\mathbf{p}_q \! = \! [\xi_{q,1}\!, \ldots\!, \xi_{q,N}]^T$}. In (\ref{EqAsymptOutputVol}), {\small$\mathbf{M}^\prime_k$} returns a {\small$N$}-by-{\small$N$} matrix whose $k$\,th diagonal is made of ones, while all the other entries are zero. Here, for $k \!> \!0$, $k$ is the index of the $k$\,th diagonal above the main diagonal (whose index $k=0$); for $k\!<\!0$, $k$ is the index of the $|k|$\,th diagonal below the main diagonal. For instance, {\small$\mathbf{M}^\prime_0 = \mathbf{I}_{N\times N}$}. Given {\small$K \! = \! 1$} and {\small$\mathbf{p}_1 = \frac{1} {\sqrt{N\Lambda_1}} \mathbf{1}_{N\times 1}$} (i.e. power is uniformly allocated across the frequency domain channels), (\ref{EqAsymptOutputVol}) becomes
\begin{equation}
\label{EqAsymptOutputVol_UniformPwrAlloc}
v_{\text{out},q}^\prime \! = \! \beta_2 E \Lambda_1 \! + \! 3\beta_4 E^2\Lambda_1^2/2 \!+\! \beta_4 E^2\Lambda_1^2 N(\!N\!-\!1)(2N\!-1)/(2N^2)\,.
\end{equation}
This equation indicates that when {\small$N$} is sufficiently large, {\small$v_{\text{out},q}^\prime$} can almost scales with {\small$N$} linearly. With the weighted-sum criterion, the asymptotically optimal waveform design problem can be formulated as
\begin{IEEEeqnarray}{cl}
\label{OriginalProblem_AsymptWsum}
\max_{\{\mathbf{p}_q\}_{q=1}^K} \quad & \left\{ \textstyle{\sum_{q=1}^K} w_q \cdot v_{\text{out},q}^\prime: \sum_{q=1}^K \Lambda_q \|\mathbf{p}_q\|^2 = 1 \right\}\,.
\end{IEEEeqnarray}
Similarly to Section \ref{SecGeneWirelessChan}, (\ref{OriginalProblem_AsymptWsum}) is reformulated as
\begin{IEEEeqnarray}{cl}
\label{EquiProblem_AsymptWsum}
&\min_{\gamma_1^\prime\!, \{\!\mathbf{p}_q\!\}_{q\!=\!1}^K\!, \{\!\mathbf{t}_q\!\}_{q\!=\!1}^K} \quad\quad  \gamma_1^\prime\IEEEyesnumber\IEEEyessubnumber\\
\text{s.t.} \quad & \textstyle{\sum_{q=1}^K} \! w_q \! \left(\!E^2 \Lambda_q^4 \mathbf{t}_q^H \!\mathbf{A}_0 \mathbf{t}_q \! - \!\beta_2 E\Lambda_q^2t_{q,0} \!\right) \! \leq \! \gamma_1^\prime, \IEEEyessubnumber \label{EqNoncovxConst_EquiProbAsymptWsum}\\
& \mathbf{p}_q^H \mathbf{M}^\prime_k \mathbf{p}_q = t_{q,k}\,, \forall q,k \IEEEyessubnumber \label{EqConst2_EquiProbAsymptWsum}\\
& \mathbf{p}_q^H \left[\mathbf{M}^\prime_k\right]^H \mathbf{p}_q = t_{q,k}^\ast\,, \forall q,k\neq 0 \IEEEyessubnumber \label{EqConst3_EquiProbAsymptWsum}\\
& \textstyle{\sum_{q=1}^K} \Lambda_q \|\mathbf{p}_q\|^2 = 1\,, \IEEEyessubnumber \label{EqConstNorml_EquiProbAsymptWsum}
\end{IEEEeqnarray}
where {\small$k \!=\! 0,\ldots,N\!-\!1$} and {\small$q = 1,\ldots, K$}. Then, the nonconvex problem (\ref{EquiProblem_AsymptWsum}) is solved by SA. To this end, similarly to (\ref{Epi_Problem_MaxVout_RankConstRelaxed}), the nonconvex constraint (\ref{EqNoncovxConst_EquiProbAsymptWsum}) is linearized by its first-order Taylor approximation. The AP of (\ref{EquiProblem_AsymptWsum}) at iteration $l$ can be formulated as
\begin{IEEEeqnarray}{l}
\label{ApproxProblem_AsymptWsum}
\min_{\gamma_1^\prime\!, \{\!\mathbf{p}_q\!\}_{q=1}^K\!, \{\!\mathbf{t}_q\!\}_{q=1}^K} \,  \gamma_1^\prime\IEEEyesnumber\IEEEyessubnumber\\
\text{s.t.} \, \sum_{q=1}^K\! w_q \! \left(\!- \!\beta_2 E\!\Lambda_q^2 t_{q,0} \!+\! E^2\!\Lambda_q^4\tilde{g}_q\!(\mathbf{t}_q; \mathbf{t}_q^{(l \! - \! 1)})  \!\right)\! -\! \gamma_1^\prime \!\leq \! 0, \IEEEyessubnumber\\
\quad \quad\text{(\ref{EqConst2_EquiProbAsymptWsum}), (\ref{EqConst3_EquiProbAsymptWsum}) and (\ref{EqConstNorml_EquiProbAsymptWsum})} \nonumber\,.
\end{IEEEeqnarray}
Recall that SCA establishes convergence by solving the convex AP (\ref{ApproxConvProblem_MaxVout}). However, the above AP is nonconvex. Fortunately, the global optimum of (\ref{ApproxProblem_AsymptWsum}) can be achieved by solving an equivalent problem.
Define {\small $\mathbf{C}_{q,1}^\prime \triangleq -\frac{\beta_2 E\Lambda_q^2+3E^2\Lambda_q^4\beta_4 t_{q,0}^{(l-1)}}{2}\mathbf{M}^\prime_0 - 3\beta_4 E^2\Lambda_q^4\sum_{k=1}^{N-1} [t_{q,k}^{(l-1)}]^\ast \mathbf{M}^\prime_k$} and {\small $\mathbf{A}^\prime_{q,1} \triangleq \mathbf{C}_{q,1}^\prime + [\mathbf{C}_{q,1}^\prime]^H$}.
Substituting (\ref{EqConst2_EquiProbAsymptWsum}) and (\ref{EqConst3_EquiProbAsymptWsum}) into (\ref{EqNoncovxConst_EquiProbAsymptWsum}), an equivalent form of (\ref{ApproxProblem_AsymptWsum}) can be finally written as
\begin{IEEEeqnarray}{cl}
\label{EquiProb_Approx_AsymptWsum}
\min_{\mathbf{\bar{p}}} \quad & \left\{ \mathbf{\bar{p}}^H \mathbf{A}^\prime_1 \mathbf{\bar{p}}: \mathbf{\bar{p}}^H \mathbf{\Lambda}\mathbf{\bar{p}} = 1 \right\}\,,
\end{IEEEeqnarray}
where {\small $\mathbf{\bar{p}}\! \triangleq\! [\mathbf{p}_1^T, \ldots, \mathbf{p}_K^T]^T$}, {\small $\mathbf{A}^\prime_1 \!\triangleq \! diag\{w_1\mathbf{A}^\prime_{1,1}, \ldots, w_K\mathbf{A}^\prime_{K,1}\}$}, and {\small$\mathbf{\Lambda}\!\triangleq \!diag\{\mathbf{\Lambda}_1, \ldots, \mathbf{\Lambda}_K$\}}, where all the main diagonal entries of the $N$-by-$N$ diagonal matrix {\small$\mathbf{\Lambda}_q$} are equal to $\Lambda_q$. Define {\small$\mathbf{U}_{\mathbf{\bar{\Lambda}}}$} as the matrix made of the eigenvectors of {\small$\mathbf{\bar{\Lambda}} \!\triangleq\! \mathbf{\Lambda}^{-1}\! \mathbf{A}^\prime_1$}. By analyzing the KKT conditions, the optimal {\small $\mathbf{\bar{p}}^\star \! = \! b \big[ \mathbf{U}_{\mathbf{\bar{\Lambda}}} \big]_\text{min}$}, where {\small$b \! = \! \big[ 1/ \big(\big[\mathbf{U}_{\mathbf{\bar{\Lambda}}} \big]_\text{min}^H \mathbf{\Lambda} \big[\mathbf{U}_{\mathbf{\bar{\Lambda}}} \big]_\text{min} \big) \big]^{1/2}$}. For {\small$\Lambda_q \!=\! \Lambda \,\, \forall q$}, {\small $\mathbf{\bar{p}}^\star \!=\! \sqrt{\frac{1}{\Lambda}} \big[ \mathbf{U}_{\mathbf{A}^\prime_1} \big]_\text{min}$}, where {\small$\mathbf{U}_{\mathbf{A}^\prime_1}$} collects the eigenvectors of {\small$\mathbf{A}^\prime_1$}. Note that the EVD of {\small$\mathbf{\bar{\Lambda}}$} (or {\small$\mathbf{A}^\prime_1$}) yields complexity of {\small$O\big((KN)^3\big)$}.

The algorithm is summarized in Algorithm \ref{AlgAsymptWsum}. As (\ref{EquiProb_Approx_AsymptWsum}) yields the global optimum of (\ref{ApproxProblem_AsymptWsum}) and {\small$\tilde{g}_q(\mathbf{t}_q^{(l)}\!; \mathbf{t}_q^{(l)})\! =\! g_q(\mathbf{t}_q^{(l)}) \!\leq\! \tilde{g}_q(\mathbf{t}_q^{(l)}\!; \mathbf{t}_q^{(l-1)})$}, the objective function of (\ref{ApproxProblem_AsymptWsum}) decreases over iterations. Then, it can be shown that the solution finally converges to a stationary point of the original problem (\ref{EquiProblem_AsymptWsum}). The detailed proof is omitted due to space constraint.

\section{Simulation Results}
\label{SecSimResults}
In the simulations, we consider a typical large open space indoor or outdoor wireless channel at a central frequency of $5.18$\,GHz with 10\,MHz bandwidth. Therefore, the channel model D \cite{MS98} for ETSI HiperLAN/2 simulation is exploited on account of i.i.d. spatial domain channel gains. The pass loss (i.e. large-scale fading) is set as $61$\,dB \cite{MBA98}, and EIRP at the BS is fixed as $36$\,dBm, i.e. {\small$PM = 3.9811$}\,W.

\begin{figure}[!t]
\centering
\includegraphics[width = 2.2in]{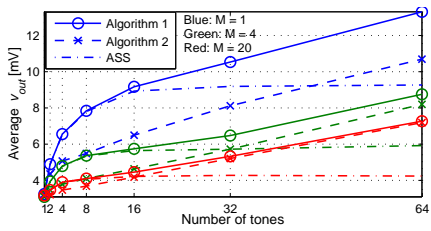}
\caption{Average $v_\text{out}$ as a function of $N$.}
\label{Fig_Comp_SCA_SA_ASS_N}
\end{figure}
Fig. \ref{Fig_Comp_SCA_SA_ASS_N} studies the average {\small$v_\text{out}$} as a function of {\small$N$}, with {\small$M \!\in\! \{1, 4, 20\}$} and  {\small$K\!=\!1$}. In the simulation, the adaptive single sinewave (ASS) scheme \cite{CBmar16arxiv} is considered as a baseline. ASS performs matched beamforming and allocates all power to the sinewave corresponding to the strongest frequency domain channel. Hence, ASS is optimal for the maximization of the 2nd order truncation model (i.e. the term containing $\beta_2$ in (\ref{EqFuncVoutq})), which essentially is the conventional linear model\cite{Wetenkamp83, ZZH13}. It is shown that given {\small$M$}, the performance gain achieved with Algorithm 1 (i.e. the SCA-based algorithm) over ASS scales with {\small$N$} and becomes significantly large. This comes from the fact that with a fixed bandwidth, as {\small$N$} increases, the frequency domain channel power gains are distributed within a narrower range. Hence, allocating all the power to the strongest frequency domain channel can be strictly suboptimal. Additionally, (\ref{EqAsymptOutputVol_UniformPwrAlloc}) implies that as {\small$N$} increases, the value of the 4th order term can be sufficiently large, such that this term may not be neglected during optimization. Therefore, it is also observed in Fig. \ref{Fig_Comp_SCA_SA_ASS_N} that when {\small$N$} is small (e.g. {\small$N=8$}), increasing {\small$M$} cannot significantly enlarge the performance gain of Algorithm 1 over ASS.

Fig. \ref{Fig_Comp_SCA_SA_ASS_N} also illustrates that given {\small$N$}, although the channel gains are not i.i.d. across frequencies, the performance gap between Algorithm 1 and Algorithm 2 (i.e. the SA-based algorithm) decreases, as {\small$M$} increases. This indicates that {\small$M$} is large enough and the channel is frequency-selective enough to make $v_\text{out}$ independent from {\small$M$}, as shown in (\ref{EqAsymptOutputVol}).

\begin{figure}[!t]
\centering
\includegraphics[width = 2.5in]{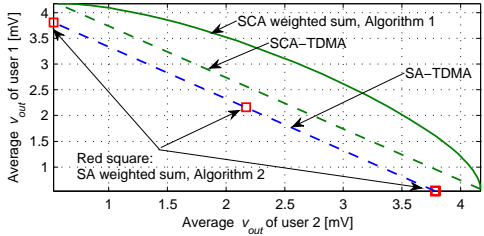}
\caption{Achievable $v_\text{out}$ region, with $M=20$ and $N=10$.}
\label{Fig_VoutRegion}
\end{figure}
Fig. \ref{Fig_VoutRegion} studies the achievable {\small$v_\text{out}$} region with {\small$K \!=\! 2$} and {\small$\Lambda_1 \! = \! \Lambda_2 \! = \! 61$}\,dB. The regions of the weighted sum algorithms are achieved by averaging {\small$v_\text{out}$} over 300 channel realizations, across various user weight pairs {\small$(w_1,w_2)$}. In SCA-TDMA (or SA-TDMA), the two users are served in a time division manner, and the optimal waveform for each user is computed by Algorithm 1 (or 2). It is shown that the achievable region of Algorithm 1 is larger than that of SCA-TDMA. That is, by generating optimal waveforms, Algorithm 1 can perform a better tradeoff between the {\small$v_\text{out}$} of the two users. As the small-scale fading CSI is not exploited in the optimization for SA-TDMA, the achievable region of SA-TDMA is significantly smaller than that of SCA-TDMA. For the same reason, Algorithm 2 is outperformed by Algorithm 1. It is also observed that Algorithm 2 only achieves three average {\small$v_\text{out}$} pairs. This is due to Algorithm 2 being only a function of {\small$\Lambda_q$} and {\small$w_q$} but not the small-scale fading channels. Further, with {\small$\Lambda_1 = \Lambda_2$}, the solution produced by Algorithm 2 only relies on {\small$(w_1,w_2)$}. Specifically, when {\small$w_1\! \neq \! w_2$}, all the power is always allocated to {\small$\mathbf{p}_q^\star$} with {\small$q^\star = \arg \max_{q} w_q$}. This is equivalent to the TDMA scenario where only one user is served. When {\small$w_1\! =\! w_2$}, all the power is randomly allocated to either {\small$\mathbf{p}_1$} or {\small$\mathbf{p}_2$}, with equal probabilities. This is equivalent to the TDMA scenario where the two users equally share the time resources.

\section{Conclusions}
\label{SecConclu}
In this paper, we have proposed efficient waveform optimization algorithms for the multiuser large-scale multi-antenna multi-sine WPT. It is shown that given a moderately large number of antennas, the low-complexity SA-based algorithm can yield solutions close to that of the SCA-based algorithm. Moreover, in the presence of a sufficiently large number of tones, the average output voltage achieved by the nonlinear-model-based waveform design can be significantly higher than that offered by the linear-model-based design. In contrast, in the presence of a small number of tones, the linear and non-linear-based designs lead to similar performance.

\bibliographystyle{IEEEtran}
\bibliography{IEEEabrv,BibPro}

\end{document}